\documentclass[conference]{IEEEtran}
\usepackage{silence}
\usepackage[style=base]{caption}
\usepackage{subcaption}
\usepackage{lmodern}
\usepackage{graphicx} 
\graphicspath{{./Images}}
\usepackage{epsf}
\usepackage{verbatim}
\usepackage{comment}
\usepackage{amssymb}
\usepackage{array}
\usepackage{setspace}
\usepackage{amsmath}
\usepackage[amsthm,thmmarks]{ntheorem}
\usepackage[ruled,lined]{algorithm2e}
\usepackage{multirow} 
\usepackage{tabularx}
\usepackage{xfrac}
\usepackage{breqn}
\usepackage{bbm}
\usepackage{enumerate}
\usepackage{algorithmic}
\usepackage{dsfont}
\usepackage{mathtools}
\usepackage[margin=1.72cm]{geometry}
\usepackage{url}
\usepackage{cite}
\usepackage{xspace}

\usepackage{color}
\makeatletter
\newcommand*{\rom}[1]{\expandafter\@slowromancap\romannumeral #1@}
\makeatother

\usepackage[labelformat=simple]{subcaption}

\newcommand\numberthis{\addtocounter{equation}{1}\tag{\theequation}}

\begin{document}
\title{\SYSFullC}
\newcommand{\SYSFullC}{Attention-based Learning for Sleep Apnea and Limb Movement Detection using Wi-Fi CSI Signals\xspace}
\newcommand{\SYSFull}{Attention-based Learning for Sleep Apnea and Limb Movement Detection using Wi-Fi CSI Signals\xspace}
\newcommand{\SYS}{IPF\xspace}

\author{\IEEEauthorblockN{Chi-Che Chang, An-Hung Hsiao, Li-Hsiang Shen, Kai-Ten Feng, Chia-Yu Chen}

\IEEEauthorblockA{
Department of Electronics and Electrical Engineering, National Yang Ming Chiao Tung University, Hsinchu, Taiwan\\
Email: chang871031.ee10@nycu.edu.tw, e.c@nycu.edu.tw, gp3xu4vu6.cm04g@nctu.edu.tw,\\ ktfeng@nycu.edu.tw, chiayu.c@nycu.edu.tw
}}

\maketitle


\begin{abstract}
Wi-Fi channel state information (CSI) has become a promising solution for non-invasive breathing and body motion monitoring during sleep.
Sleep disorders of apnea and periodic limb movement disorder (PLMD) are often unconscious and fatal.
The existing researches detect abnormal sleep disorders in impractically controlled environments. 
Moreover, it leads to compelling challenges to classify complex macro- and micro-scales of sleep movements as well as entangled similar waveforms of cases of apnea and PLMD.
In this paper, we propose the attention-based learning for sleep apnea and limb movement detection (ALESAL) system that can jointly detect sleep apnea and PLMD under different sleep postures across a variety of patients.
ALESAL contains antenna-pair and time attention mechanisms for mitigating the impact of modest antenna pairs and emphasizing the duration of interest, respectively.
Performance results show that our proposed ALESAL system can achieve a weighted F1-score of 84.33, outperforming the other existing non-attention based methods of support vector machine and deep multilayer perceptron.

\end{abstract}




\section{Introduction}
Wireless sensing technologies have brought great attention in diverse fields, such as factory automation, home security, and human-centric sensing\cite{6G},\cite{csiratio}.
With increasing concern about personal health, non-invasive healthcare is considered one of the important techniques for observing physical conditions.
Several researchers have developed through ultra-wideband (UWB) radars \cite{UWBradar} and Wi-Fi signals \cite{physical}.
Nevertheless, UWB radar requires additional dedicated hardware, which is not widely deployed in existing environments.
By contrast, pervasive Wi-Fi devices are available in most environments, which rely on received signal strength (RSS) and channel state information (CSI).
However, RSS is a coarse information possessing unified pathloss values from multipath, which is easily influenced by shadowing effect and noise providing ambiguous data \cite{RSSref}.
Fine-grained CSI can contribute informative signal features for detecting imperceptible motions, such as breath or limb movement\cite{tl}.
For breath detection, sleep apnea is a common and severe breathing disorder that causes the body to decrease the rate or even stop breathing during sleep.
In addition, periodic limb movement disorder (PLMD) is a symptom potentially accompanied by sleep apnea with brief and repetitive twitching or kicking of the extremities during sleep \cite{sleepstandard3}.
Both obstructive sleep apnea (OSA) and PLMD induce a period of unstable variations in CSI signals.
Moreover, the degree of freedom of these movements varies in different stature, making it difficult to distinguish between apnea and PLMD.

Currently, there are numerous studies researching on estimation of breathing and body motion based on CSI signals \cite{resilient,SixTypes,apnea1,apnea2,biRNN,RMD,seizure}.
In \cite{resilient}, the amplitude and phase of CSI are jointly employed for respiratory rate monitoring.
The authors in \cite{SixTypes} use the amplitude response to train a deep multilayer perceptron (DMLP) for distinguishing different respiratory patterns.
In works of \cite{apnea1} and \cite{apnea2}, apnea detection is conducted by adopting the threshold-based methods of the periodicity-level and peak detection, respectively.
In sleep movement recognition, paper\cite{biRNN} utilizes a bidirectional recurrent neural network (RNN) for extracting temporal features to recognize different types of body movement during sleep.
\cite{RMD} adopts support vector machine (SVM) to classify distinct rhythmic movement disorder.
The authors in \cite{seizure} develop a mathematical model to empirically classify normal body movements and nocturnal seizures.
However, existing studies monitor abnormal breathing and apnea under a controlled environment with fixed postures, which is impractical for OSA.
Moreover, large-scale body movements are experimented and easily to be detected due to obvious dissimilarity.

To the best of our knowledge, this is the first work to consider micro-scale signals entangled from apnea and PLMD syndrome.
Conventional threshold-based methods and machine learning models lead to suboptimal results.
Since the threshold values and models need to be specifically adjusted for different people.
This provokes a challenging task, which should be designed by advanced deep learning techniques incorporating available signal domains.
The main contributions of this work are summarized as follows.

\begin{itemize}
    \item [1)]
We propose the attention-based learning for sleep apnea and limb movement detection (ALESAL) system. 
To the best of our knowledge, this is the first work to jointly consider sleep apnea and PLMD detection using CSI signals with open medical data from diverse patients under arbitrary sleep postures.
\end{itemize}

\begin{itemize}
    \item [2)]
ALESAL system contains several attention mechanisms for detecting under uncontrolled scenarios.
We design the time attention neural network to emphasize the duration of interest due to time-varying spectrograms revealing breathing rate.
Since different antenna pairs provide different quality of time-domain features, antenna-pair attention is performed to mitigate the impact of modest antenna pairs.
\end{itemize}

\begin{itemize}
    \item [3)]
We have evaluated our ALESAL system between different deep learning architectures and other existing methods.
Benefiting from joint time and antenna-pair attention mechanisms, the proposed ALESAL system can achieve the highest weighted F1-score of 84.33 compared to other existing methods.
\end{itemize}

The rest of this paper is organized as follows. Section II describes the system architecture. Section III introduces the proposed ALESAL system. Section IV provides the performance comparison. Finally, the conclusions are described in Section V.
\section{System Architecture}
\begin{figure} 
    \centering 
    \setlength\belowcaptionskip{-1.3\baselineskip} 
    \includegraphics[width=0.36\textwidth] {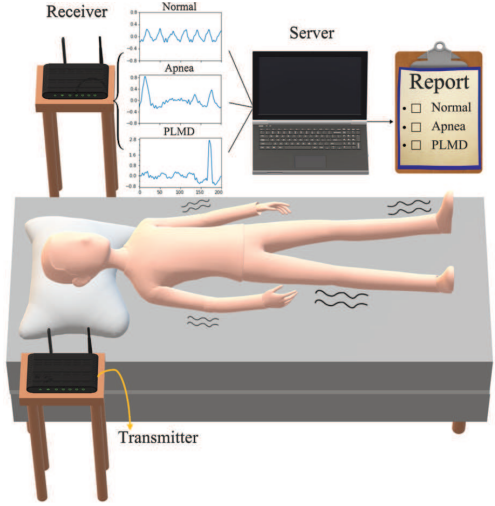} 
    \caption{System architecture for contactless sleep apnea and PLMD detection using commercial Wi-Fi CSI signals.}
    \label{Images:場景圖3_1}
\end{figure}
In this section, we introduce the system architecture of Wi-Fi monitoring sleep apnea and PLMD using CSI signals.
As shown in Fig. \ref{Images:場景圖3_1}, the scenario is composed of two Wi-Fi access points, one as the transmitter and the other one as the receiver.
Both are operated under multiple-input multiple-output (MIMO) system.
Furthermore, orthogonal frequency division multiplexing (OFDM) is considered for data transmission, i.e., the total spectrum is divided into multiple orthogonal subcarriers.
With the benefit of MIMO and OFDM, CSI reflects the channel status over all subcarriers, which characterizes how signals propagate from the transmitter to the receiver under the effects of scattering, fading, and power decay.
We have a total of $S$ subcarriers and $P$ antenna pairs.
The received signal at the receiver can be modeled as
\begin{align}
    \ y_{i,p}\left(t\right) = h_{i,p}\left(t\right)x_{i,p}\left(t\right)+n_{i,p}\left(t\right),
\end{align}
where $y_{i,p}\left(t\right)\in\mathbbm{C}$ and $x_{i,p}\left(t\right)\in\mathbbm{C}$ represent the received and transmitted complex signals with the $i^{th}$ subcarrier of the $p^{th}$ antenna pair at the $t^{th}$ time instant, respectively.
$h_{i,p}\left(t\right)\in\mathbbm{C}$ and $n_{i,p}\left(t\right)\in\mathbbm{C}$ are the CSI and additive white Gaussian noise (AWGN), respectively.

Typically, the indoor Wi-Fi signals propagate with the direct path and the reflection paths affected by the human body, walls, and furniture.
As a result, CSI at the receiver is the superposition of all multipath components, which can be written as
\begin{align*}
\begin{footnotesize}
\begin{aligned} 
    h_{i,p}\left(t\right)\!=\! \sum_{l_{p}=1}^{L_{p}}A_{l_{p}}\left(t\right)\exp\left(-j2\pi\frac{d_{l_{p}}\left(t\right)}{\lambda}\right)\!=\!\left|h_{i,p}\left(t\right)\right|\exp\left(j\angle h_{i,p}\left(t\right)\right),
\end{aligned}
\end{footnotesize}
\numberthis
\end{align*}
where $l_{p} \in \left[1,L_{p}\right]$ is the index of multipath components from the $p^{th}$ antenna pair with $L_p$ paths.
$A_{l_{p}}\left(t\right)$ and $d_{l_{p}}\left(t\right)$ are the complex attenuation and propagation distance of the ${l_{p}}^{th}$ path, respectively. $\lambda$ is the wavelength of the wireless signal.
$\left|h_{i,p}\left(t\right)\right|$ denotes the absolute value of amplitude response, whilst $\angle h_{i,p}\left(t\right)$ stands for the phase response.
Since the propagation distance of the direct and reflection paths from static objects are generically time-invariant, these CSI signals may remain unchanged.
However, human motions during sleep alter the multipath, such as subtle chest movement or limb jerking, which results in fluctuation of the received CSI.
These effects are captured in the measured CSI that can be applied to monitor the patient's sleep conditions.
In addition, due to the regular up-down tempo at the chest during normal breathing, signal paths that reflected off the chest vary periodically as a sinusoidal-like waveform.
Owing to the uncontrollable brief twitching or kicking of the extremities during sleep, the amplitude response has an abrupt change that disrupts the periodic waveform associated with respiration.

\section{Proposed Attention-based Learning for Sleep Apnea and Limb Movement Detection (ALESAL) System}
\begin{figure} 
  \centering
  \setlength\belowcaptionskip{-1.2\baselineskip}
  \includegraphics[width=0.48\textwidth]{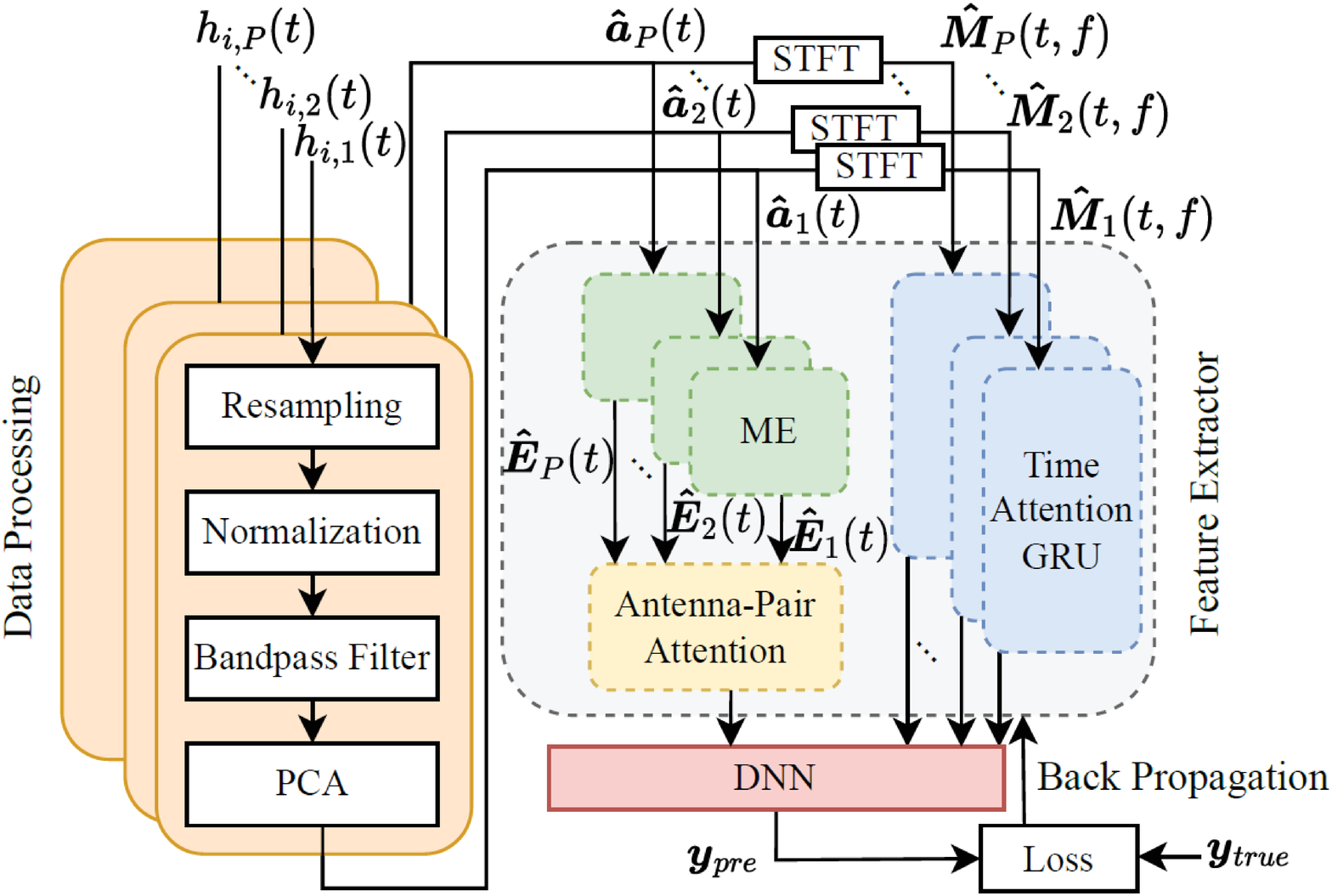}
  \caption{Schematic diagram of the proposed ALESAL system.}
  \label{Images:架構圖_3}
\end{figure}
In ALESAL, we have designed a system based on attention mechanisms for sleep apnea and limb movement detection.
As illustrated in Fig. \ref{Images:架構圖_3}, we first obtain CSI data from Wi-Fi and then apply data preprocessing to remove environmental noise.
The CSI amplitude response fluctuates with the chest and limb movements.
The shape of the waveform contains information, such as breathing depth and rhythm, which reveals the patient’s condition.
In addition, short-time Fourier transform (STFT) is leveraged to acquire the frequency-domain information.
The spectrogram captures the breathing signal changes over time.
Both the time- and frequency-domain information are performed as the inputs of the feature extraction module, including morphology extractor (ME) modules collaborated with the antenna-pair and time attention gated recurrent unit (GRU) modules.
The detailed process of each module is described as follows.

\subsection{Data Preprocessing}
Suffering from packet loss and transmission delay, the receiver is unable to receive every packet with the same frequency.
To accurately label the data from ground truth, the raw CSI data is resampled to cope with missing information.
Considering the uncontrollable power variations from the defects of commercialized devices, we apply normalization to each antenna pair using the maximum and minimum values of amplitude response among all subcarriers within an antenna pair to remove the effects, which can be expressed as
\begin{align}
    \hat{h}_{i,p}\left(t\right) = \frac{\left|\tilde{h}_{i,p}\left(t\right)\right| - \mathop{\min}\limits_{ \forall q } \left|\tilde{h}_{q,p}\left(t\right)\right|}{\mathop{\max}\limits_{ \forall q } \left|\tilde{h}_{q,p}\left(t\right)\right| - \mathop{\min}\limits_{ \forall q } \left|\tilde{h}_{q,p}\left(t\right)\right|}, \forall q \in\left[1,2,...,S\right],
\end{align}
where $\left|\tilde{h}_{i,p}\left(t\right)\right|$ represents the resampled amplitude response of the $i^{th}$ subcarrier with a fixed sampling rate $f_{r}$.
Afterwards, we can obtain the normalized CSI time series in a time window from the $p^{th}$ antenna pair, which can be denoted as $\boldsymbol{\hat{h}}_{i,p}\left(t\right) = \left[\hat{h}_{i,p}\left(t\right)\right],\quad\forall t \in \left[t-Z,t-Z+1,...,t\right]$, where $Z$ is the size of the time window.

The signals reflected off the body encompass human motions from the chest and limbs and random noise.
It is essential to alleviate the noise from the received signals for further processing.
Therefore, we apply a Butterworth bandpass filter to every subcarrier to eliminate high-frequency noise, which can be represented as
\begin{align}
B\left(f\right)=\left\{
\begin{array}{rcl}
1,       &      & { b_{min} \leq f \leq b_{max}, }\\
0,     &      & {\text{otherwise}.}\\
\end{array} \right. \label{bandpass}
\end{align}
The filtered signals are represented as $\boldsymbol{\hat{c}}_{i,p}\left(t\right) = \boldsymbol{\hat{h}}_{i,p}\left(t\right) \ast \hat{B}\left(t\right)$, where $\hat{B}\left(t\right)$ is the inverse fast Fourier transform (IFFT) of $B\left(f\right)$ in \eqref{bandpass} and $\ast$ indicates the convolution operator.

Due to OFDM and MIMO, we can acquire high-dimensional CSI signals from diverse subcarriers and multiple antenna pairs.
Owing to high computational cost, we adopt principal component analysis (PCA)\cite{drunkdriving} to reduce the data dimension.
Note that in each antenna pair, we extract the first principal component that preserves most information across all subcarriers.
The first principal component of the $p^{th}$ antenna pair can be expressed as
\begin{align}
    \boldsymbol{\hat{a}}_{p}\left(t\right) = PCA\left(\left[\boldsymbol{\hat{c}}_{1,p}\left(t\right),...,\boldsymbol{\hat{c}}_{S,p}\left(t\right)\right]\right),
\end{align}
where $PCA\left(\cdot\right)$ indicates the function that converts a $S$-dimensional signal to a 1-dimensional signal, i.e., the original $Z\times S$ filtered signal is projected to $Z\times 1$ dimension signal.
Additionally, we utilize STFT to transform the time-domain signal $\boldsymbol{\hat{a}}_{p}\left(t\right)$ into the spectrogram to capture the breathing signal changes over time, which can be expressed as
\begin{align}
    \boldsymbol{\hat{M}}_{p}\left(t,f\right)=\int_{t+\tau}^{t-\tau} w\left(t-\tau\right) \boldsymbol{\hat{a}}_{p}\left(t\right)\exp\left(-j2 \pi f \tau\right) d\tau, \label{stft}
\end{align}
where $\boldsymbol{\hat{M}}_{p}\left(t,f\right)$, $w\left(t-\tau\right)$ and $\tau$ are the spectrogram, Hann window function, and window size, respectively.

\begin{figure}
  \centering
  \setlength\belowcaptionskip{-1.3\baselineskip}
  \includegraphics[width=0.4\textwidth]{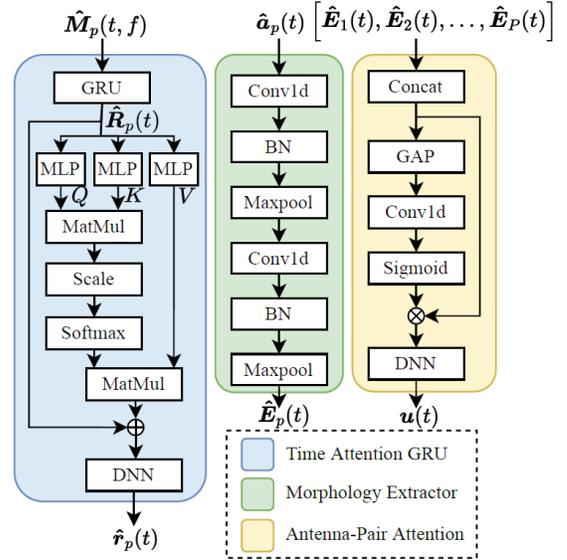}
  \caption{Neural network model for ALESAL. "MatMul" denotes inner product of two matrices. "BN" stands for batch normalization. "GAP" and "Conv1d" respectively indicate the operation of global average pooling and 1D convolution.}
  \label{fig:network}
\end{figure}
\subsection{Time Attention GRU}
Compared with traditional RNN, long short term memory (LSTM) and GRU can capture the feature of time and prevent the gradient vanishing problem by using the gate mechanism.
In addition, GRU has fewer gates than that in LSTM, which leads to fewer parameters and lower complexity.
The time attention GRU module is proposed to identify the key segments of a time series, which contains information that can discriminate sleep events.
For this purpose, our system adopts a self-attention layer with a residual neural network to cooperate with the GRU, as illustrated in Fig. \ref{fig:network}.

The extracted temporal features from the spectrogram can be computed as
\begin{align}
\boldsymbol{\hat{R}}_{p}\left(t\right) = GRU_{p}\left(\boldsymbol{\hat{M}}_{p}\left(t,f\right);\boldsymbol{w}^{g}_{p},\boldsymbol{b}^{g}_{p}\right),
\end{align}
where ${GRU_p}\left(\cdot\right)$ indicates the operation of temporal feature extraction by GRU layer from the $p^{th}$ antenna pair. 
$\boldsymbol{w}^{g}_{p}$ and $\boldsymbol{b}^{g}_{p}$ denote the weights and bias of GRU layer, respectively.
The GRU network will learn to control the reset and update gates to retain information that needs to be remembered for a certain period.
For example, when the patient has brief unconscious limb jerking, the information of frequency components attributed to limb movement will be remembered due to its distinctive components different from apnea and normal breathing.

Moreover, the self-attention mechanism is adopted to calculate the correlations between each time interval, which strengthens the significant intervals with larger weights.
As illustrated in Fig. \ref{fig:network}, we employ a self-attention layer with a residual network connection, which is represented as
\begin{align}
    \boldsymbol{\hat{r}}_{p}\left(t\right) = \bigg(softmax\left(\frac{ \boldsymbol{Q}\boldsymbol{K}^{T}}{ \sqrt{d_{k}} }\right)\boldsymbol{V} + \boldsymbol{\hat{R}}_{p}\left(t\right)\bigg) \cdot \boldsymbol{w}^{D}_{p}+\boldsymbol{b}^{D}_{p}, \label{time_attention_equation}
\end{align}
First,  we obtain the Query ($\boldsymbol{Q}$), Key ($\boldsymbol{K}$), and Value ($\boldsymbol{V}$) through three distinct linear transformations on the GRU output feature $\boldsymbol{\hat{R}}_{p}\left(t\right)$.
Then, the similarity between $\boldsymbol{Q}$ and $\boldsymbol{K}$ is calculated by inner product and normalized by the square root of the dimension of $\boldsymbol{K}$ denoted as $d_k$.
$softmax\left(\cdot\right)$ indicates a softmax-based activation function.
Besides, in order to avoid the loss of original features and gradient vanishing problem, a residual connection is applied to strengthen the original features.
Finally, we obtain the latent features $\boldsymbol{\hat{r}}_{p}\left(t\right)$ through a linear layer, where ${\boldsymbol{w}_{p}^{D}}$ and ${\boldsymbol{b}_{p}^{D}}$ respectively denote the weights and bias.

\subsection{Morphology Extractor}
The CSI amplitude response fluctuates with the chest and limb movements.
Accordingly, the shape of the amplitude waveform contains information, such as breathing depth and rhythm, which reveals the patient's condition.
As shown in Fig. \ref{fig:network}, we adopt a convolution-based approach in the morphology extractor (ME) module to capture the latent feature from CSI as
\begin{align}
\boldsymbol{\hat{E}}_{p}\left(t\right) = CNN_{p}\left(\boldsymbol{\hat{a}}_p\left(t\right)\right),
\end{align}
where ${CNN_{p}}\left(\cdot\right)$ and $\boldsymbol{\hat{E}}_{p}\left(t\right)$ indicate the feature extraction function and output feature vectors of the ME module from the $p^{th}$ antenna pair, respectively.
Note that the batch normalization, activation function of rectified linear units, and max pooling are included in a generic CNN architecture.
We then concatenate the output feature maps from all $P$ morphology extractor modules, which is the input of antenna-pair attention module.
\subsection{Antenna-Pair Attention Module}
Since there is no restriction on the sleeping position of patients, signals may not reach the receiver.
Therefore, some receiving antennas are not capable to collect complete signals, causing potentially modest antenna pairs with insufficient information about human motions.
In order to mitigate the impact of modest antenna pairs, we explore the ECANet\cite{ECANET} structure to let the model adaptively pay attention to informative antenna pairs, as shown in Fig. \ref{fig:network}.
The output feature maps of the ME modules are concatenated as
\begin{align}
    \boldsymbol{\tilde{E}}\left(t\right) = \left[\boldsymbol{\hat{E}}_1\left(t\right),\boldsymbol{\hat{E}}_2\left(t\right),...,\boldsymbol{\hat{E}}_P\left(t\right)\right].
\end{align}
Then, the global average pooling (GAP) operation is performed on the concatenated feature, given by
\begin{align}
\boldsymbol{\tilde{e}}\left(t\right)\!=\!g_{n}\left(\boldsymbol{\tilde{E}}\left(t\right)\right)\!=\!\frac{1}{L}\sum_{l=1}^{L}\hat{e}_n\left(l\right), \forall n \in \left[1,2,...,N\right],
\end{align}
where $L$ is the length of $\boldsymbol{\hat{e}}_{n}$, which is the $n^{th}$ CNN output feature vector from $\boldsymbol{\tilde{E}}\left(t\right)$.
$N$ is the number of CNN output features from previous ME modules.
Next, a fast 1D convolution operation followed by a Sigmoid activation function is utilized to learn the correlations between CNN output features, which refers to the attention weights
\begin{equation}
    \boldsymbol{w} = \sigma\bigg(C1D_{k}\left(\boldsymbol{\tilde{e}}\left(t\right)\right)\bigg),
\end{equation}
where $\boldsymbol{w}$ represents the attention weights, a relatively high value denotes the antenna pair with more information about human motions, while a lower value denotes the antenna pairs with insufficient information.
$C1D\left(\cdot\right)$ and $\sigma\left(\cdot\right)$ indicate the 1D convolution operation and Sigmoid activation function, respectively.
$k$ is the kernel size of 1D convolution, which is nonlinearly correlated with the number of CNN output features\cite{ECANET}
\begin{align}
    k = \left|\frac{\log_2\left(N\right)}{\gamma}+\frac{b}{\gamma}\right|_{odd},
\end{align}
where $\gamma$ and $b$ are positive controlling constants.
$\left|x\right|_{odd}$ indicates the nearest odd number of $x$.
Afterwards, output $\boldsymbol{u}(t)$ can be acquired by multiplying the obtained weight $\boldsymbol{w}$ and convoluted concatenated feature $\tilde{\boldsymbol{E}}(t)$ and passing through a linear layer, which is represented as
\begin{align}
    \boldsymbol{u}\left(t\right) = \left(\boldsymbol{\tilde{E}}\left(t\right) \cdot \boldsymbol{w}\right) \cdot \boldsymbol{w}^{d}+\boldsymbol{b}^{d},
\end{align}
where $\boldsymbol{w}^{d}$ and $\boldsymbol{b}^{d}$ indicate the weights and bias of the linear layer, respectively.
The output vector $\boldsymbol{y}_{pre}$ can be acquired from the DNN layer, leveraging features from the time attention GRU and the antenna-pair attention modules as input.
Finally, the model parameters are updated by the backpropagation mechanism to minimize the typical cross-entropy loss calculated by the ground truth label $\boldsymbol{y}_{true}$ and the probability vector $\boldsymbol{y}_{pre}$, i.e., $L_{CE} =-\frac{1}{N_t}\sum_{j=1}^{N_t}\boldsymbol{y}_{true}\left(j\right)^T\log\boldsymbol{y}_{pre}\left(j\right)$, where $N_t$ and $T$ respectively denote the total number of training data and operation of transpose.

\section{Performance Comparison}
\begin{table}[t]
    \begin{center}
    \caption {Parameter Settings}
        \begin{tabular}{ll}
            \hline
            System Parameter & Value \\ \hline \hline  
            The number of antenna pairs $P$ & 4\\
            The number of subcarriers $S$ & 114\\
            Frequency band $f_c$ & 5 GHz\\
            Sampling rate $f_r$ & 10 Hz \\
            Window size $Z$ & 20 seconds \\
            $[b_{min}\;b_{max}]$ for band-pass filter & [0.1 2] Hz\\
            Time window of STFT $\tau$ & 10 seconds \\ 
            Number of CNN output feature $N$ & 64\\
            Length of CNN output feature vector $L$ & 22\\ 
            Kernel size of antenna-pair attention $k$ & 3\\
            Parameters for antenna-pair attention $\gamma, b$ &    $2$, $1$ \\
            \hline
        \end{tabular} \label{tab:table1}
    \end{center}
    \end{table}
In performance comparison, we evaluate the proposed ALESAL system for joint sleep apnea and PLMD detection under uncontrolled environments with arbitrary postures.
We utilize open medical data recorded from various patients.
The work in \cite{opendata} has provided extensive real-world patient studies at a sleep clinic with four contactless devices for estimating respiratory rate.
They observed twenty patients during their overnight sleep while employing wearable sensors to record the ground truth labels of respiratory rate.
We utilize Wi-Fi CSI open dataset in \cite{opendata}, where two commercial Wi-Fi devices are operating at 5 GHz frequency band with Atheros chip.
Each device has two antennas forming a MIMO-OFDM system with 114 subcarriers per antenna pair.
The system parameters are summarized in Table \ref{tab:table1}.
Empirically, we set $k$ as $3$ for the highest accuracy and choose some patients as the training set and some as the testing set.
We conduct training by utilizing data from a total of 6 patients including three classes, normal breathing, apnea, and PLMD. 
The training set has sizes of $\{4744, 3531, 4104\}$ for each class.
In the training set, $10\%$ of data is used for validation to avoid overfitting.
The testing set is with sizes of $\{1353, 627, 1016\}$ for each class from other $5$ patients.

\begin{figure}[t]
  \centering
  \begin{subfigure}[b]{0.49\textwidth}
    \includegraphics[width=\textwidth]{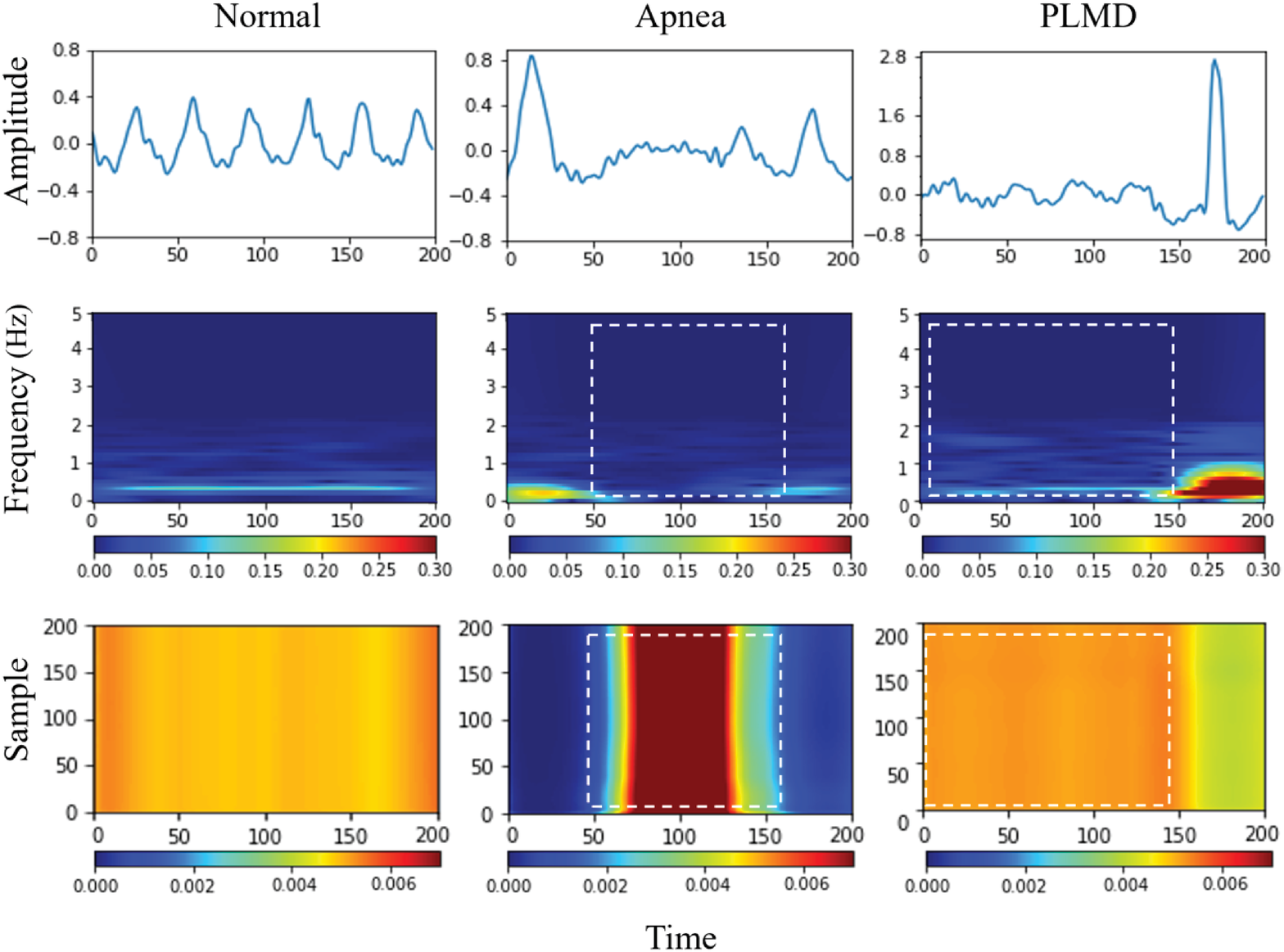}
    \subcaption{}
    \label{fig:Time Attention}
  \end{subfigure}  
  \begin{subfigure}[b] {0.49\textwidth}
    \includegraphics[width=\textwidth]{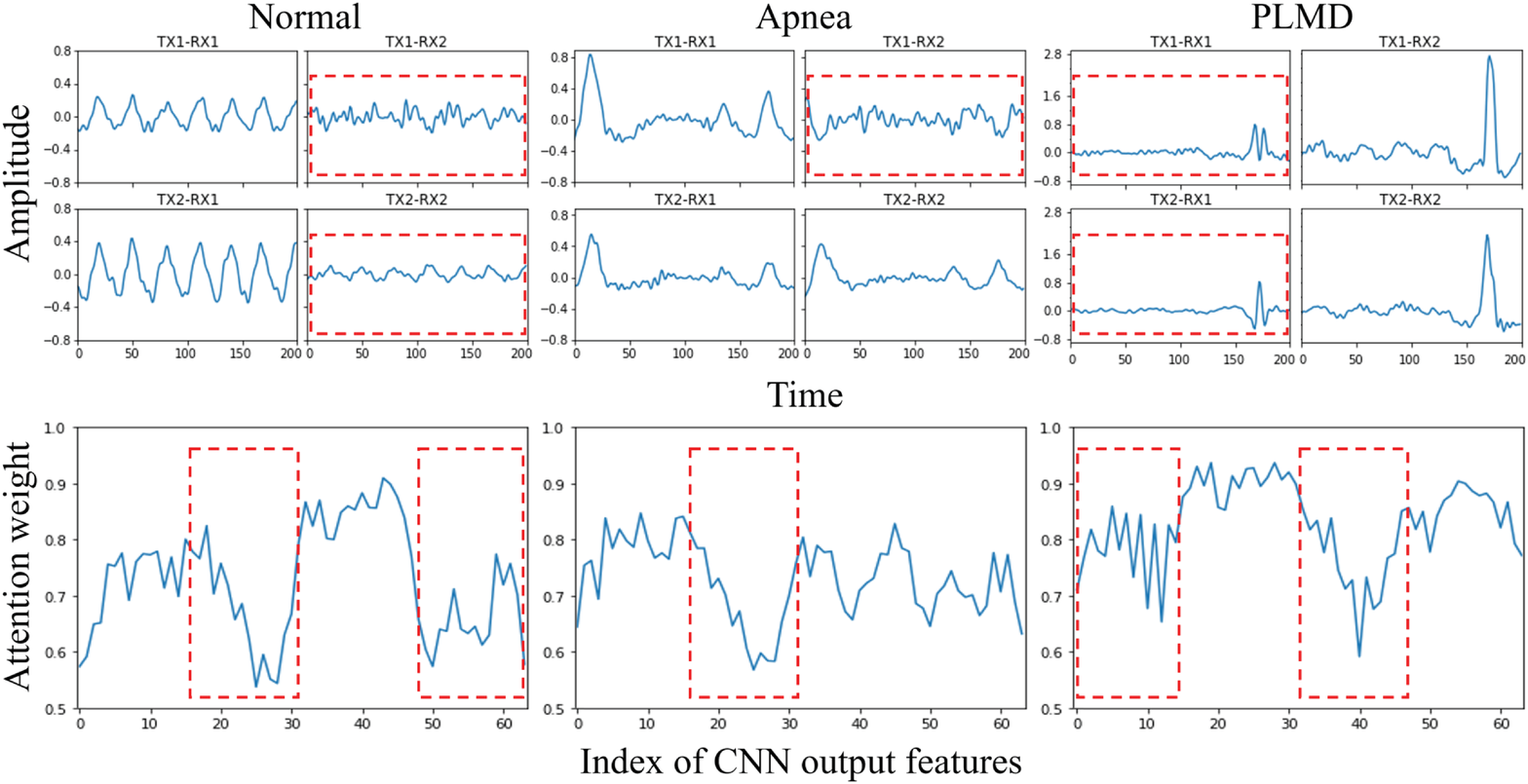}
    \subcaption{}
    \label{fig:Pair_Attention_2}
  \end{subfigure}
  \setlength\belowcaptionskip{-1.3\baselineskip}
\caption{The observation of attention weights of (a) time attention in GRU and (b) antenna-pair attention (Note that indexes of attention weights from 1-16 stands for pair 1, 17-32 for pair 2, 33-48 for pair 3 and 49-64 for pair 4).}
\label{fig:layout}
\end{figure}

\begin{figure}[t]
  \centering
  \begin{subfigure}[b]{0.24\textwidth}
    \includegraphics[width=\textwidth]{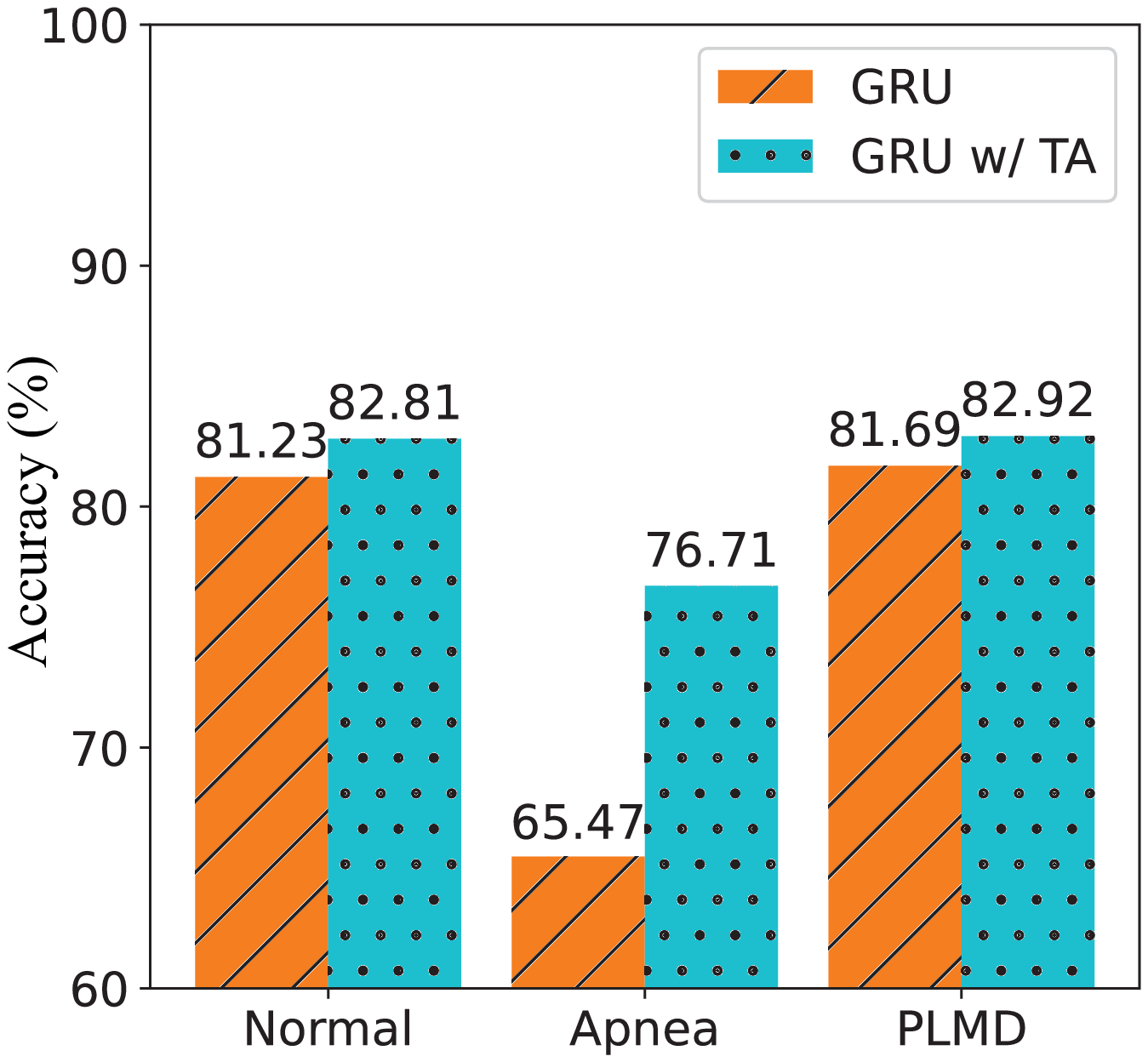}
    \subcaption{}
    \label{fig:time attention comparision}
  \end{subfigure}  
  \begin{subfigure}[b] {0.24\textwidth}
    \includegraphics[width=\textwidth]{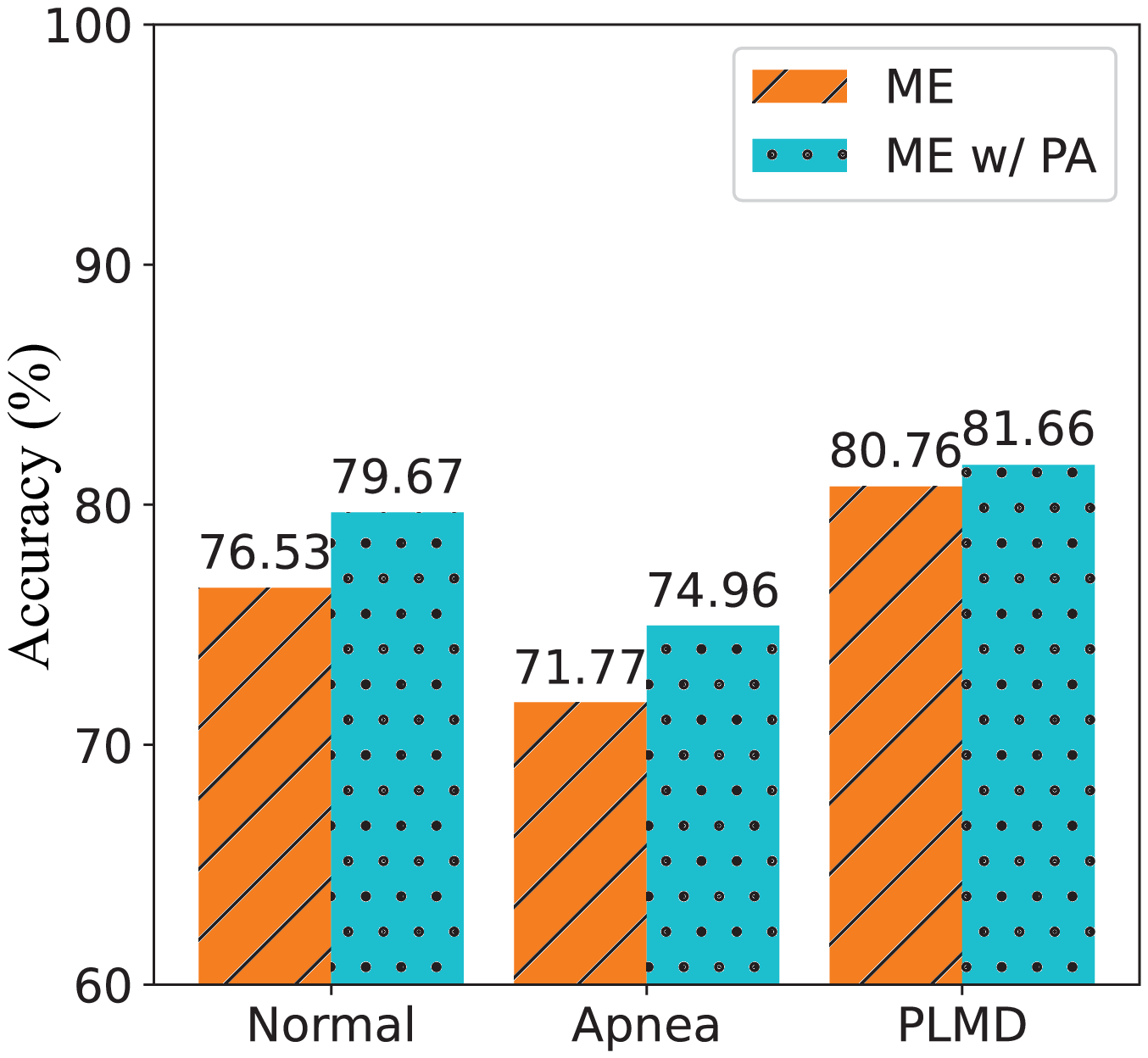}
    \subcaption{}
    \label{fig:pair attention comparision}
  \end{subfigure}
  \setlength\belowcaptionskip{-1.3\baselineskip}
\caption{Accuracy performance of ALESAL for (a) GRU with/without time attention and (b) ME with/without antenna-pair attention. "TA" denotes time attention, whereas "PA" indicates antenna-pair attention.}
\label{fig:interior_results}
\end{figure}

In Fig. \ref{fig:Time Attention}, we reveal the attention weights in the time attention GRU module.
Each column represents individual classes of normal, apnea, and PLMD.
The respective rows of Fig. \ref{fig:Time Attention} illustrate the normalized amplitude waveform from a single antenna pair, spectrum $\boldsymbol{\hat{M}}_{p}\left(t,f\right)$ of respiratory rate, and attention weights.
As shown in cases of apnea and PLMD, the white dashed lines indicate that the model will pay more attention to the time intervals with the abnormal situations.
Fig. \ref{fig:Pair_Attention_2} depicts the attention weights in the antenna-pair attention module.
The upper four subplots illustrate the normalized amplitude waveform from four antenna pairs ($2$ transmitters and $2$ receivers), while the lower plot demonstrates the attention weights.
Because sleeping towards different orientations will produce distinct reflective paths, we can observe that signals received by some antenna pairs are comparatively ambiguous, as circled by red dashed boxes.
We can observe the modest antenna pairs are given smaller attention weights.
Therefore, as shown in Fig. \ref{fig:interior_results}, we acquire higher improvements by employing either time or antenna-pair attention in the apnea class than the others due to the larger difference in attention weights, i.e., accuracy improvement of around $11\%$ for GRU w/ TA and of about $3.2\%$ for ME w/ PA. 
To elaborate a little further, little improvement of around $1\%$ is observed due to irregular and entangled waveforms of PLMD.

\begin{figure}
  \centering
  \setlength\belowcaptionskip{-1.3\baselineskip}
  \includegraphics[width=0.37\textwidth]{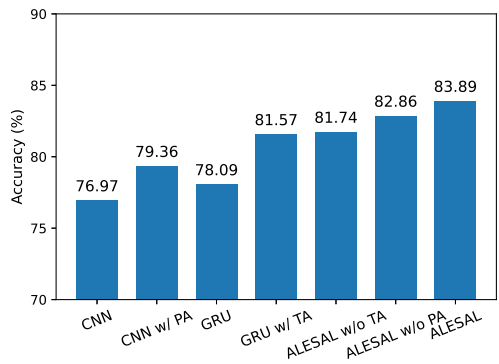}
  \caption{Performance comparison between proposed ALESAL system and the deep learning-based methods.}
  \label{fig:Comparison_other}
\end{figure}
As illustrated in Fig. \ref{fig:Comparison_other}, we compare the proposed ALESAL system with other deep learning methods for distinguishing sleep events.
We perform the same data preprocessing method for each training method.
The first two bars are for conducting ME module with only amplitude input.
The third and fourth bars apply only GRU modules with only spectrum features.
For the fifth and sixth bars, we neglect the time attention and antenna-pair attention module in ALESAL, respectively.
It can be observed that the proposed ALESAL system outperforms the other baselines with the highest accuracy of 83.89\% since our system benefits from both time and frequency features with respective attention mechanisms.

\begin{figure}
  \centering
  \setlength\belowcaptionskip{-1.3\baselineskip}
  \includegraphics[width=0.42\textwidth]{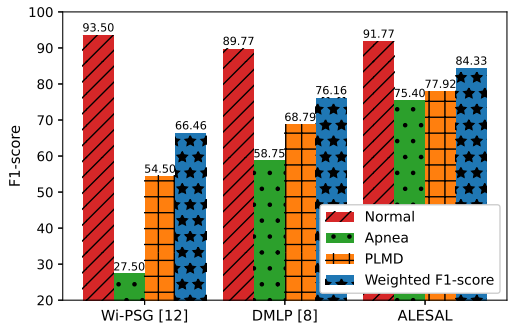}
  \caption{Performance comparison between ALESAL and other existing methods in open literature.}
  \label{fig:exterior_comparison}
\end{figure}

Fig. \ref{fig:exterior_comparison} shows the performance comparison between our ALESAL and the existing methods in open literature.
The Wi-PSG \cite{RMD} calculated different statistics from CSI data to train SVM classifier for differentiating disorders.
For a fair comparison, we use our data preprocessing scheme and extract statistics from four different antenna pairs to train SVM classifiers and average the results of four classifiers.
The DMLP network in \cite{SixTypes} is trained by the amplitude response of CSI and applied to classify breathing patterns.
As shown in Fig. \ref{fig:exterior_comparison}, the proposed ALESAL scheme outperforms the other existing methods with the highest weighted F1-score of 84.33 for sleep apnea and PLMD detection under uncontrolled scenarios.

\section{Conclusion}
This paper has proposed an ALESAL system that can detect apnea and PLMD syndrome in the open medical dataset from real-world patients.
In our proposed system, joint time and antenna-pair attention are designed to leverage frequency- and time-domain features.
In time attention, we employ self-attention with a residual neural network to emphasize the significant duration. 
For antenna-pair attention, the model pays attention to effective latent features to mitigate modest antenna pairs.
Performance comparisons have demonstrated that our proposed system benefited from both attention mechanisms acquiring a higher accuracy.
Compared with existing methods, ALESAL can achieve the highest F1-score of 84.33 in sleep apnea and PLMD detection in the open medical dataset consisting of diverse patients.

\footnotesize
\bibliographystyle{IEEEtran}
\bibliography{VTC_2022_12_12}    


\end{document}